# Interactive Visual Exploration of Halos in Large Scale Cosmology Simulation

Guihua Shan, Maojin Xie, Feng'An Li, Yang Gao, Xuebin Chi

**ABSTRACT**

Halo is one of the most important basic elements in cosmology simulation, which merges from small clumps to ever larger objects. The processes of halos' birth and merging play a fundamental role in studying the large scale cosmological structure's evolution. In this paper, a visual analysis system is developed to interactively identify and explore the evolution histories of thousands of halos. In this system, an intelligent structure-aware selection method in What You See Is What You Get manner is designed to efficiently define user's interesting region in 3D space with 2D hand-drawn lasso input. Then the exact information of halos within this 3D region is identified by data mining in the merger tree files. To avoid visual clutter, all the halos are projected in 2D space with a MDS method. Through the linked view of 3D View and 2D graph, Users can interactively explore these halos, including the tracing path and evolution history tree.

## 1. INTRODUCTION

In cosmology, dark matter halo is a cluster of dark matter whose density exceeds the threshold. During large-scale cosmology simulation process, even-distributed dark matter tends to group together under gravity and forms small halos. Small halos collide and merge with each other and form ever larger halos. Halo is one of the most fundamental elements in large scale simulation. In N-body simulation, halos can be found with Friend-Of-Friend (FOF) method (Riebe et al. 2011). A merger tree is a tree-like structure which records a halo's merging history. It's a special tree. A tree node may contain several halos and the most massive halo in the node is named as the master halo while the others are called satellite halos. At some point, a merger tree records the local evolution process of the universe, which lays an important position for merger tree.

Large-scale cosmology simulation generates large numbers of halos, which range from thousands to millions. Large numbers of halos and their merger trees make things much sophisticated and bring great challenges in halo exploring and merger history back tracing.

It is impossible to show all of the tracing paths or merger trees together in one time for all particles or halos, so it is important to interactively explore the interesting objects and study the physical processes and selection of interesting particles from visualization context is essential for interaction. The selected interesting area can be further explored, which not only help the user focus on the interesting object but also dramatically reduce the processing data volume to make further exploration more efficient. For example, astronomer prefers to select an interesting halo structure from current view to trace back its evolution or the related statistic information.

Although there are lots of visualization works for cosmological data, most of them focus on rendering of particles data(Hassan et al.2011, Lipsa et al.2012), where the interactive functions are limited to real-time rendering and no further information feeds back except the continuous changes of the scene.

Manually selection on the rendering image with 2D mouse allows users to obtain a spatial area by multiple operations. But it's hard for people to define an exact 3D surface to encircle the desired spatial region with 2D input, which is tedious and time consuming. Moreover, when an interesting halo is found, they may want to know not only its properties like mass, radius and velocity, but also its evolution history or the merging location both in space and time. A more intelligent and efficient interaction with large scale temporal dark matter data is needed.

In this paper, we propose an interactive visual analysis system. The workflow of the system is illustrated in Figure2. A WYSIWYG particle selection method is provided, in which Cloud-in-Cell (CIC) method is adopted to generate grid data and then marching cubes are employed to gain isosurface and use max projected area as criteria to automatically filter the user's most wanted particle cluster. With our method, a 3D region is easily to define with a hand-drawn 2D lasso, which makes the selection process of halos in 3D space quite efficient. And then we propose our visual analysis scheme. Several 2D/3D views are organized as linked view style and halo information is shown in different perspective. 2D layout for halos makes halo size clear. Multidimensional Scaling (MDS) is employed in 2D layout to avoid overlapping and its distance-reserved feature makes the halos' relationship more clear. We use 2D and 3D layout for merger trees. Dark matter origin and sub-halos moving path are more clear in 3D layout while merge time is easier to perceive in 2D layout. The main contributions of the paper are:

a) An intelligent structure-aware selection method in WYSIWYG manner is designed, which provide an intuitive and efficient way to interact with large scale data.

---

G. Shan, M. Xie, F. Li, Y. Gao, X. Chi
SuperComputing Center, Computer Network Information Center, Chinese Academy of Sciences

{sgh,xiemaojin,lifa,gaoyang,chi}@sccas.cn

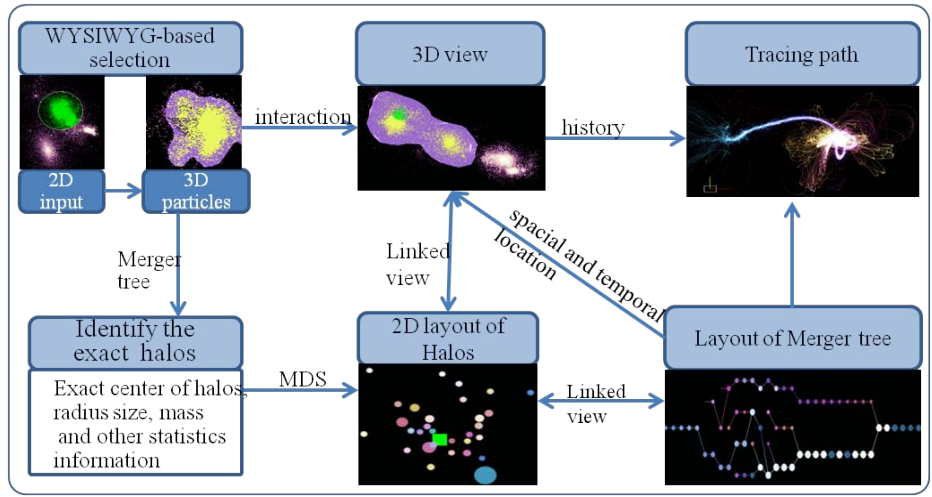

Fig.1 The workflow of visual analysis system

b) 2D halo layout with MDS project avoids the visual clutter, which allows an efficient choose for the interesting halo.

c) With the interaction in three space(3D view, 2D projected graph and merge tree), the system provide a visual analysis for deep exploration of halos.In the remainder of the paper, firstly we review the related work in Section 2. Then we describe the visual analysis techniques of the WYSISWG selection algorithm in Section 3 and halo exploration method in Section 4. Finally we make a conclusion of the paper in Section 5.

## 2. RELATED WORK

Visualization and visual analysis, which extract and present useful information out of petabytes of cosmology data, are essential for astronomy discovery. There are many such works in the past decade both in dark matter visualization, evolution history visualization and object selection.

### 2.1 DARK MATTER VISUALIZATION

Particle based simulation data usually employ some kinds of splatting approaches and approximate the spatial distributions of physical quantities using kernel interpolation techniques. Interactive large-scale dark matter visualization is proposed by Fraedrich et al. (2009), which employs hierarchy quantization and data compression to reduce data volume. GPU acceleration is used for decompress on the fly. Interactive visualization of time-varying data set is proposed by Hopf et al.(2003), which adopted out-of-core for data accessing and B-splines for position interpolation.

Except splatting approaches, volume rendering based dark matter visualization is also proposed by Fraedrich et al.(2010). Particle sets are represented in a level-of-detail manner and perspective grid is employed to allow effectively reducing the amount of primitives to be processed in run time. Kaehler et al.(2012) propose a novel approach for cosmic web visualization which take use of full phase-space information to generate a tetrahedral tessellation of the computational domain, with mesh vertices defined by the simulation's dark matter particle position.

As Hassan et al. (2011) and Lipsa et al.(2012) mention respectively that many visualization works for cosmology simulation focused on single time step, while temporal data are usually visualized in batch mode. Although some interactive visualization works has been proposed as mentioned previous in this session, most interactive works are restricted to real-time rendering, with no more information feeding back except the continuous changes of the scene. This is of vital importance in interactive exploration, but is still far out of the satisfaction for scientists to explore more detail information and make discovery when they are warding in the scene, for example, when an interesting halo is found, they may want to know not only its properties like mass, radius and velocity, but also its evolution history or something other.

### 2.2 EVOLUTION HISTORY EXPLORATION

Di et al(2008) visualize black holds with arrows indicating their masses and location in multiple constant images. Takle et al.(2012) perform tracking and building halo evolution history with merger tree in parallel and show the merge tree in 2D. Fluke et.at (2009) develop S2PLOT, a novel multipurpose visualize tool for interactively visualize cosmology data, including 3D merger tree. By adopting VRML, they are able to interchange the interactive 3D scientific visualization among a variety of mediums. The structure is quite clear as the redshift is high but overlapping severely as redshift decrease. There are lack of an intuitive global view to select a target halo at redshift z=0, which is the most important snapshot for the domain scientists as it is the present cosmic world.

In exploratory visualization, linked view (Roberts 2004) is efficient as there are more than one variables and analysis aspects at one time. Each of variables or aspects shows in one view makes things clearer, and the linked view makes the interaction responded in all views.

### 2.3 SELECTION OF INTERESTING OBJECTS

For more intuitively and efficiently selecting the interesting area in 3D data, many approaches are developed such as ray casting based methods (Peng et al. 2010; Kohlmann et al. 2012), surface based methods (Argelaguet 2009). While Guo et al. (2011) firstly present a novel "WYSIWYG" idea for volume rendering. With intuitive user interface like Photoshop instead of traditional transfer function edition, users can easily achieve satisfied volume rendering results. Later, more work in WYSIWYG style is provided, Guo et al.(2013) describe a local WYSIWYG volume rendering by constructing contour tree in preprocess, where more useful gestures are also designed for easily controlling the volume rendering. Wiebel et al. (2012) present a novel raycasting based method, which allow user to intuitively select spatial position in volumetric renderings. Observable structures are characterized by large jumps in the accumulated opacity; the picked structure corresponds to the largest jump of the accumulated opacity. Yu et al.(2012) describe a lasso-like technique, where the grid is constructed in the box of selected area and marching cubes are used for enclose the particle cloud structure intuitively lasso on the interesting 2d project.

We also propose a WYSIWYG technology for the particle cloud selection in our system. Compared to (Yu et al. 2012), our method is able to separate the different enclosed structures of selected spatial particles, and the structure which is perceived most is chose by default. While in (Yu et al. 2012), the selected enclosed structures are unable to be separated, which is inconvenient for further analysis. Compared to (Wiebel et al. 2012), the data type and the principle to realize the "what you see is what you get/pick" is different. In our research, we work on particle cloud data instead of volume data. We use marching cubes and flood fill to get the interesting structure, instead of volume rendering and opacity.

### 3. THE WYSIWYG SELECTION ALGORITHM

The WYSIWYG particle cloud selection algorithm can automatically select the particle cluster user interested according to the drawn region. The input of the algorithm is all the particles and their attribute data. Our algorithm consists of the following five steps:

1. Preliminarily marking the particles that are projected in the user-drawn region.
2. Gridding the marked particles, and getting a regular volume data.
3. Extracting particles in clusters with larger density through the marching cubes algorithm.
4. Splitting the clusters into independent ones by an improved flood fill algorithm.
5. Obtaining the particle cluster insterested by computing and comparing each cluster's projective area. By default, we suppose that the point cluster with the largest projective area is the most interesting one. Figure 2 shows the processing flow chart of the algorithm.

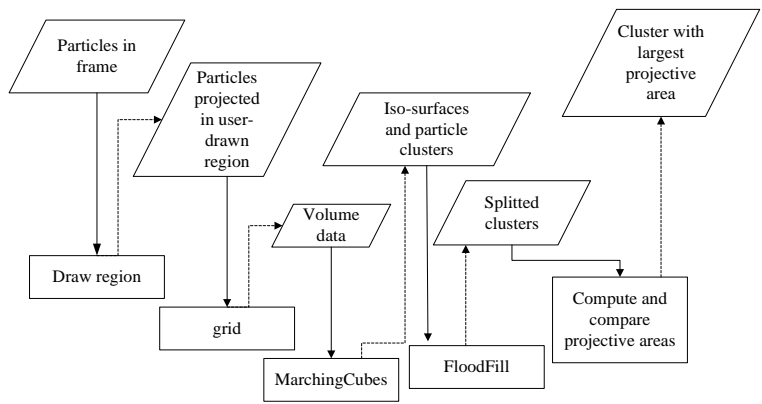

Fig.2 The processing flow chart of the WYSIWYG algorithm

### 3.1 Marking the particles in the user-drawn region

We provide two kinds of drawn region: one is circle, the other is polygon. According to the drawn lasso, we generate a filled polygon, and render it to the frame buffer. We name the rendered image a mask. Using the mask, we project all the particles in this view, and mark the particles in the drawn region. The following algorithm steps are all based on the marked particles. Figure 3 shows two kinds of user-drawn lasso.

### 3.2 Constructing grid for marked particles

We use the isosurface extraction algorithm marching cubes to select out the particle clusters which have dense distribution. Before executing the marching cubes algorithm, we construct a regular grid of the marked particles and compute the density of each node.

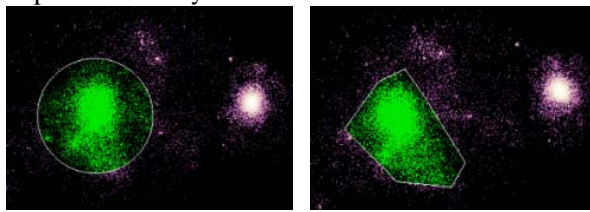

(a) Circle selection　　　(b) Polygon selection

Fig.3. Two kinds of user-drawn lasso

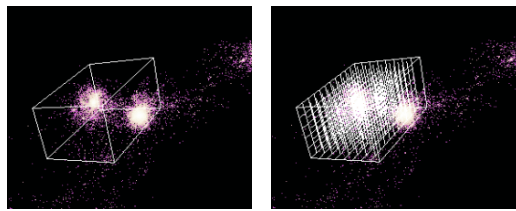

(a)The axis aligned bounding box　　(b) Volume data as a scalar field of the volume data. According to the range of the marked particles' coordinate, we obtain an axis aligned bounding box, as shown in Figure 4(a). We construct a volume data and split it into $n_x \times n_y \times n_z$ cells as Figure 5(b) shows.

Fig. 4 Voxelization of the selected region

In order to apply marching cubes to extract isosurface, we calculate particle density $D(x,y,z)$ at each node $(x,y,z)$ as the scalar filed. We calculate it as:

$$D(x,y,z) = \sum_{i=0}^{m}(d_{ix} + d_{iy} + d_{iz}) \tag{1}$$

Where $(x,y,z)$ is a grid node; m is the total number of particles inside the 8 cells which take $(x,y,z)$ as their vertex; $d_{ix}, d_{iy}, d_{iz}$ represent the contribution of each particle $i$ inside the 8 cells to the node $(x,y,z)$ in $x,y,z$ axis respectively. That is, the density of each grid node is the sum of the contribution from all particles in the 8 cells that the node belongs to. Actually, we calculate each particle's contribution to the 8 vertices of which cell that the particle belongs to. By accumulating the contribution at each vertex, we get the density of each node.

If the coordinate of the particle p is $(p_x, p_y, p_z)$, the boundary of the volume data in $x,y,z$ axis is $(\min_x, \max_x), (\min_y, \max_y), (\min_z, \max_z)$, the length of each cell is $l_{cell}$, then the relative coordinate of p for each direction $k = x,y,z$ can be calculated as

$$r_k = (p_k - \min_k)/l_{cell} \tag{2}$$

Rounding up and down to $r_k$, we obtain two nodes coordinate $k_1$ and $k_2$ closest to particle p in direction $k$:

$$\begin{cases} k1 = 0, k2 = 1; (r_k = 0) \\ k1 = floor(r_k), k2 = ceil(r_k); (0 < r_k < N_k + 1); \\ k1 = N_k; k2 = N_k + 1; (r_k = N_k + 1) \end{cases} \tag{3}$$

Where $N_k$ is the cells number in axis $k$.

According to the distance between the particle and the node, we can calculate the contribution of the particle to its adjacent grid nodes in axis $k$:

$$w_{k1} = k_2 - r_k, w_{k2} = r_k - k_1 \tag{4}$$

The contribution of particle p to the 8 grid nodes of the cell where particle p is inside can be expressed as follows:

$$contri(x_o, y_p, z_q) += w_{xo} + w_{yp} + w_{zq}, o = \{1,2\}, p = \{1,2\}, q = \{1,2\} \tag{5}$$

Density value for each node is accumulated with contribution of all particles in the user-drawn space.

### 3.3 Surface Extraction based on Marching Cubes

The application of surface extraction algorithm can divide the particles into clusters according to the distribution intensity. First, we calculate the average density $\rho_0$ of the grid nodes as the initial threshold, which could be adjusted in real time during visualization. Applying the marching cubes algorithm to the grid, we seek out regions with density $\rho \geq \rho_0$. The result of marching cubes algorithm is surfaces with density equal to $\rho_0$, and the purpose of this article is to select particles. Therefore, we need to mark the particles inside the isosurfaces. The marching cubes algorithm is processed in voxel. According to the comparison result between the density of each voxel's 8 vertices and the threshold, we classify the voxels as voxels inside the isosurface, voxels outside the isosurface and voxels on the isosurface. All of particles in the inner voxels must be located inside the isosurface; particles in the outer voxels must all be located outside the isosurface; particles in the boundary voxels require part in and part out of the isosurface.

$$CellValue_i \begin{cases} = 0, outerVoxel \\ = [1,254], boundaryVoxel \\ = 255, InnerVoxel \end{cases} \tag{6}$$

By tagging the voxels with the tags above, the isosurfaces and their inner particles can be easily mapped to the volume data, and then we can extract the particle clusters. In order to extract particles in the volume, we need to build a lookup table which records the mapping relation between the voxel and the particles inside. This lookup table can be made at the same time of grid density calculation. For inner voxels, we mark all the particles according to the lookup table. However, the boundary voxels requires further judgment for particles inside. We calculate density for each particle inside the boundary voxels. Comparing this density with the threshold, we distinguish whether it is inside the isosurface or not. Like calculating density for grid nodes, we should determine the voxel index the particle belonging to according to the particle's coordinate. The particle's density can be obtained as cubic interpolation of the voxel's 8 vertices'. If the particle density value is greater than the threshold, the particle is accounted as inside the isosurface and marked;

otherwise, it is outside and then abandoned.

## 3.4 Clusters splitting

Now we have obtained all the clusters with density greater than the threshold. In order to separate the clusters, we extend the flood fill algorithm from 2D to 3D to get the connected components through filling area. And finally, we have the clusters independent. The flood fill algorithm is commonly used to obtaining connected components in 2D images. However, this algorithm requires seed to be assigned. In this paper, we do the filling in voxels. We choose the first voxel in the isosurfaces as the first seed. In the process of filling, we mark the visited voxels, to ensure that the filling process will not revisit voxels.

The conditions to be filled go as follows:
1) The current voxel's coordinates are within the legal scope, i.e.:

$$x \in (0, N_x), y \in (0, N_y), x \in (0, N_z) \tag{7}$$

2) The current voxel has not been visited;
3) The current voxel is located inside or on the boundary of the isosurfaces.

When the voxel satisfies the above three conditions at the same time, it will be marked with the ClusterId (a variable of identity of the cluster). For the particles inside the voxel, they will be all marked with the same cluster id if the voxel is an inner voxel; or only the particles whose density is greater than the threshold will be marked with cluster id when the voxel is a boundary voxel. Considering the six-neighborhood of the seed voxel in filling processing, if at least one vertex of the voxel has density greater than the threshold, we fill the voxel and mark the particles following the principle above. After finishing one connected space, we will find the next seed voxel for next fool filling, which is marked as ClusterId+1. And repeat the above operation, until all voxels are marked as visited. Finally we will obtain the segmentation.

However when two different clusters are too near, in the case that their boundary are in two adjacent voxels or even in a same voxel, which is illustrated in Figure 5, they will be mistakenly marked as one cluster. In Figure 6, the purple parts are the isosurfaces, the yellow and green parts mark the particles in the isosurfaces. Figure 6(a) shows only yellow part which means there is one cluster in the scene. Obviously, there are two clusters actually. So filling conditions is not right for this case.

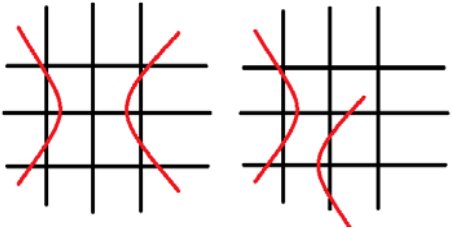
Fig.5 Two cases for wrong segmentation

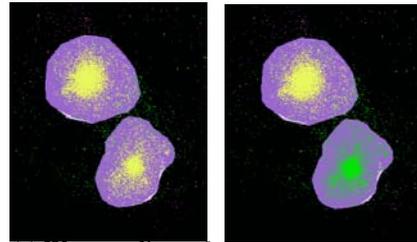
(a) Wrong split result    (b) Right split
Fig. 6 Split results in different filling conditions

We improve the filling conditions in fool fill algorithm, when marching toward each direction, we do the following judgment:

The next voxel has at least one side in the marching direction satisfying the condition that both vertices' grid node density is greater than the threshold. For example, the marching direction is x+1, and the next voxel with its vertices tagged as Figure7. The four edges in the marching direction x+1 are: e0、e1、e2、e3. If only density of vertex 3 and vertex 2 is greater than threshold, then the voxel will be filled.

If the condition above is true, then fill the next voxel and selectively mark the particles inside.

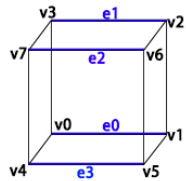

Fig. 7 A voxel with its vertices tagged

When comparing the grid node density of each voxel with the threshold in the marching cubes algorithm, we record the comparison result between both vertices' density of voxel's four edges with threshold in x/y/z directions as Tag. So Tag is an integer range from 0 to 7. Each bit of Tag stores two marching directions' result. At least one edge satisfies the condition that both vertices' density is greater than the threshold, and then this bit is signed as 1. While, if all the 4 edges in the marching direction can't satisfy the condition, this bit is signed as 0. When do the filling algorithm, we make a decision to continue or stop depending on the Tag value. With this improved algorithm, we get the correction segmentation in Figure 6(b).

## 3.5 Hardware accelerated computation of projected areas

We have separated the different particle groups in section 3.4. For scientists, it is preferable to automatically select out the particle cluster that they are most interested in. We assume that the cluster occupies the largest projection area in the user-drawn region is the one scientists most wanted. This, of course, is true in most cases since they really perceive the largest projection area when they make a lasso in the rendering image.

Therefore, the default output of the selected cluster, should be the one with the largest projection area. In order to calculate each cluster's area, the vertices data of each isosurface is required. So not only the particles inside should be split, the isosurface encircled each cluster should also be stored respectively. At the time we do the marching cubes for each voxel, one lookup table recording the voxel id and vertices information of the isosurfaces in it is built, as shown in Figure9 (a). When splitting the clusters using flood fill algorithm, store the isosurfaces' vertices according to the clusters respectively, and then we get another lookup table as Figure 9 (b) shows.

(a) Voxel id and vertex sequence

(b) Cluster id and vertex sequence

Fig. 9 Lookup tables

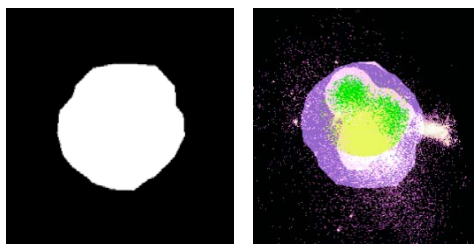
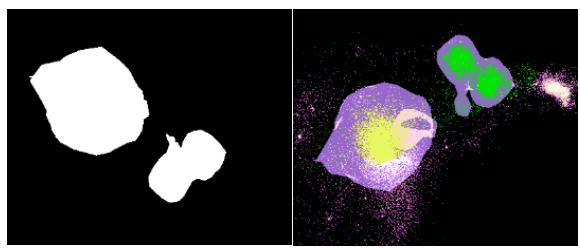

(a) The overlapped projection  (b) The isosurface sence

Fig.10 The projection and the isosurface scene in original view positions

(a) The overlapped projection  (b) The isosurface sence

Fig 11 The projection and the isosurface scene in different view

Using the ClusterId as index of vertex sequence of isosurfaces, we render every encircled isosurface to the frame buffer. Figure 10(a) shows the isosurface generated by user selection in the current view direction. Figure 10(b) shows the original isosurfaces and particles in the same view direction. Since we render all isosurfaces in the same view in their original positions, the result projection is the overlapped area. Changing the viewpoint, we get the separated projection as shown in Figure 11(a). Projecting the isosurfaces according to their locations is not desirable, because there are two isosurfaces in this case in fact.

In order to avoid overlapping of each isosurface's projection, we render them into N viewports, where N is the amount of clusters. We split the window along x axis into N viewports and render each isosurface in one viewport. The resulting projection area is squeezed in the x direction, but all projection areas are squeezed with the same ratio, thus there is no influence in the following area comparison step.

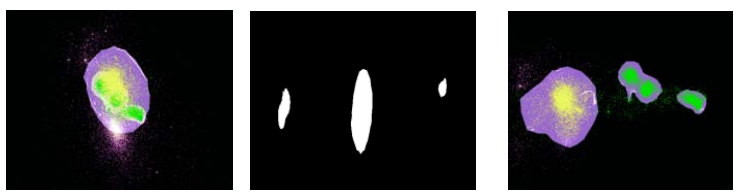

Fig. 12 The projection and original isosurfaces

By counting the non-black pixels of each projection, we get the largest cluster and return its ClusterId to the user. Finally, we mark the cluster returned in yellow, as shown in Figure 12. The two green clusters are independent and can also be extracted by this method.

## 4 HALO EXPLORATION

With WYSIWYG technology, we are able to extract interesting halos efficiently from a chaotic data and analyze its evolution process in linked view style. We will describe our technology implementation and case studies in this chapter.

We performed the experiments on a machine with an Intel® Core™ i5-3470 3.20GHz CPU, 8GB RAM and an NVidia Quadro 600 graphics card with 1GB video memory. The data is from the cosmology dark matter simulation with n-body method using 1 million particles. The output is a scatter point dataset with 64 time steps, where each particle has 6 variables such as ID, position, velocity, mass, dispersion, and density. The number of halos in all time steps add up to 536048.

### 4.1 HALO LAYOUT AND INTERACTION

Technically, we build a lookup table which records the mapping relationship between the voxels and the halos inside with the similar method in section3.3. The different is that the center positions of master halos substitute for particle positions.

Halos displayed in 3D space may suffer from overlapping and occlusion, which makes interactive analysis not so convenient. To get a better layout for master halos in 2D view, we employ MDS to map halos' positions from 3D space to 2D space, while halos' distances are reserved as much as possible. Thus the halos belonging to the same FOF group will cluster together. We use a disc to represent the halo, with its radius mapping to halo's radius, and its color and brightness mapping to velocity dispersion and density respectively. Thus we can perceive the main properties of each halo in an intuitive way.

When interacting with a halo, we click on that disc representing it in 2D view. And then the halo id is determined by comparing halo center positions and the cursor position under Manhattan distance.

### 4.2 CASE STUDY1-TRACING PATH

In this case study we focus on how to show the particles evolution of the universe in an intuitive way. It is one of the most important problems that astronomers concerned. Scientists used to make animation to study the evolution. However, since human's memory is temporary, it is easy to lose detail information when the observation is mutative. Another problem bothered scientists when analyzing evolution of the universe is that visualizing the time sequence of all data may bring heavy IO burden. Low efficiency has always been the main obstacle for further research in cosmogony. All the problems mentioned above can be solved by using our system. Scientists can select out the most interesting particles by the 2D input lasso. Then we draw an image of the evolution trace path of the selected particles. We use the "cyan-blue-purple-yellow" color table to represent time sequences from far to near on the time line (see Figure 13 for example). The image alone contains the evolution process of the selected particles. Compared with the animation, our method is more intuitive, and need less IO. In addition to the space trace path, we can clearly see the dark halo merger, divergence and other physical processes in the development of structure.

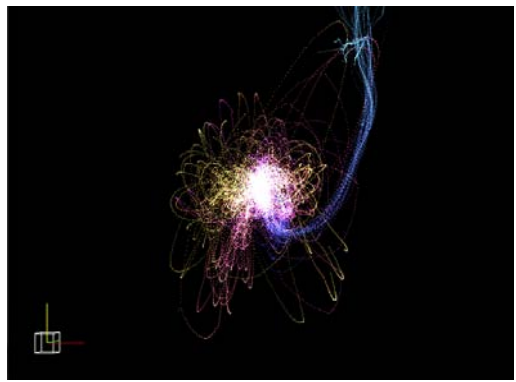

Fig.13 An image of particle cloud trace path

### 4.3 CASE STUDY 2- EVOLUTION MERGER TREE OF A HALO

Although trace path is a nice way to express the evolution process in 3D space, the merger tree is a better quantization expression when exploring the merger history, both in mass and time.

We draw the merge tree in a level-wise way, with the level representing time. Disc is used to express the halo, with its size corresponding to halo size, while density and velocity dispersion are encoded into brightness and color. Thus we can easily perceive the main attributes of each halo in the merger tree. Edges with gradually varied color are used to express the merger relationship.

When we select a region in the 3D view, we can retrieve a set of halos. They are scatter in 2D region, with their positions transform by MDS, while the relative distances are reserved. We can see from Figure 14(a) that after two small clusters of particles selected (shown in yellow), halos inside are show as discs in 2D in top right view, with two main clusters, respectively. We can click on the discs, they will be highlighted with another color and its merger tree will display in bottom view. From the merger tree, we can clearly find two main streams (mark as 1 and 2 with red dash circles, we call it stream 1 and stream 2 respectively) merged into the main stream of selected halo.

We can also easily notice that there are no merging after stream1 and stream2 merging, but the density and velocity dispersion change from high to lower, which means that the halo is collided and some of the mass is split away. In Figure 14 (b), the different interested cluster is selected, the tracing path a selected root halo is showed in 3D view instead of one time step space view, which illustrate intuitive of the halo evolution in the 3D spatial and temporal axes.

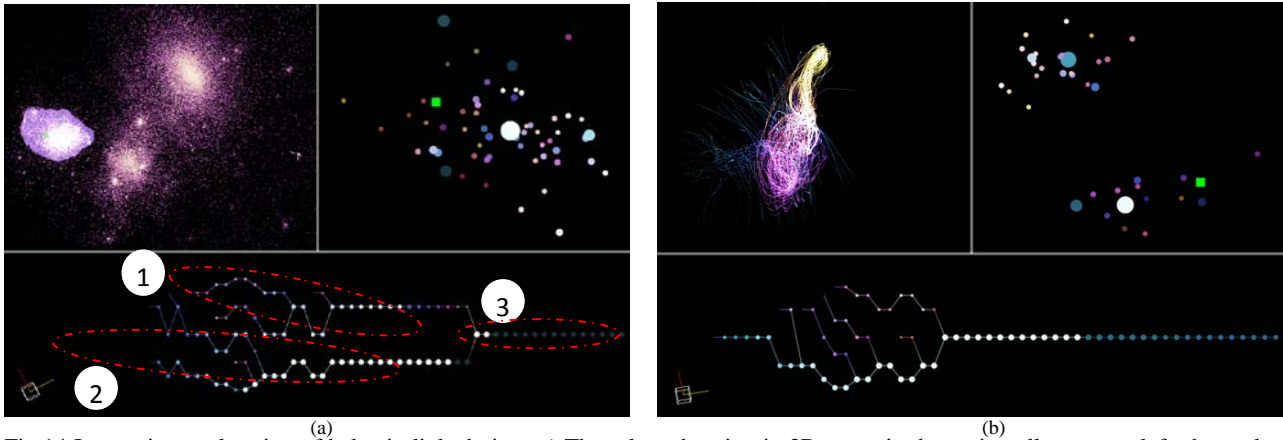

(a) (b)

Fig.14 Interactive exploration of halos in linked view. a).The selected region in 3D space is shown in yellow at top left, then selected halo is shown in green both in 2D( top right) and 3D view(top left), the merger tree of selected halo is shown in bottom view. From the merger tree, we can clearly find two main streams (mark as 1 and 2 with red dash circles, we call it stream 1 and stream 2 respectively) merged into the main stream of selected halo, there are no merging after stream1 and stream2 merging, but the density and velocity dispersion change from high to lower b) The different interesting region is selected. the selected halo in 2D space is shown in green at top right. the evolution path of this halo is showed in 3D View (top left). The merger tree is shown in the bottom.

The performance when dealing with the data which has 1 million particles and 536048 halos is shown in the Table 1. In interactive exploration, only in the first stage all the particles are rendering for a globe view. The small interesting number of halos is load for temporal analysis after the user-drawn selection. The total rendering time includes particles rendering, MDS projection of the halos in the isosurface, merger subtree searching for the root halo and rendering of the subtree. It ranges from 0.15s to 0.3s depending on the total number of halos in the subtree. If the selected root halo is merged by more halos, then it takes more time when searching and rendering. From Table 1, we also can see that the merger tree process is time consuming. It takes about 80% of the total time. This is because when doing the merger tree pass, we need to retrieval the halo data sequence to find out the parent halo recursively. The WYSIWYG selection method takes 0.03s in average, but this process is execute only at the beginning and not recomputed if the different root halo is selected in the MDS projection graph, so this selection method consumes little time.

Tabel 1. Performance (Time in seconds)

| The number of halos in merger subtree of the selected root halo | 7873 | 2083 | 188 | 46 |
|---|---|---|---|---|
| Merger subtree searching and rendering time | 0.25 | 0.18 | 0.15 | 0.14 |
| Total time | 0.33 | 0.26 | 0.19 | 0.17 |

## 5 CONCLUSION

We designed a cosmology visualization system for halo analysis. An efficient WYSIWYG selection scheme is proposed and halos can be easily selected for further analysis. We adopted MDS in transforming halos' positions from 3D space to 2D space in order to reserve the relative distance and get an intuitive 2D layout. The 3D space, 2D graph and 2D merge tree are combined together to support an interactive exploration of halos evolution. In our case study, we analysis the halo's evolution process in two ways, one is trace path, with moving path clearly shown in 3D view, another is merger tree, with the merger process clearly shown in quantities, both in time and mass.

**Acknowledgements:** The authors are grateful to Prof. Feng Longlong from Purple Mountain Observatory, CAS and Doctor Zhu Weishan from Nanjing Univ. for the data set and instructive suggestions. The work is supported by Chinese National Sciences Foundation No. 91230115.